# Symmetry breaking and effective photon mass

Yong-Hae Ko, Gwang-Il Kim, Sok-Hyon Won and Nam-Chol Kim[*]

*Faculty of Physics,* **Kim Il Sung** University*, Taesong District, Pyongyang, DPR of Korea*
[*]ryongnam19@yahoo.com,

**Abstract**

We suggest a possibility that the photon can acquire a finite mass in a medium when the external interaction and symmetry is broken on the basis of Chern-Simons gauge theory.

1. **Introduction**

The possibility that the photon acquires a finite mass has been discussed theoretically [1-3] and investigated experimentally by several researchers.

The acquisition of the photon mass due to the breakdown of Lorentz symmetry by the external magnetic field [1] and that of the chiral symmetry [2] have been developed. And the effects on the cosmic microwave background (CMB) due to the photon mass have been expected [4, 5].

Many direct and indirect tests for the photon mass have been performed under several physical conditions and the search for upper limit on the photon mass has increased over the past several decades [6-8]. The effects of the Aharonov-Bohm type of light in moving media were discussed and used to determine the limit of the photon mass [9]. In addition, as light bending is confirmed with growing accuracy, it provides a way of setting upper bounds on the photon mass, especially by means of a Very Long Baseline Interferometry (VLBI) [10]. In recent papers, the upper limits of photon mass is regarded to be less than $10^{-65}\,g$ [11].

In this paper, we suggest a possibility that the photon can acquire an effective mass in a medium due to the electromagnetic interaction.

Considering the electromagnetic interaction in a medium according to the quantum field theory, mediating particles of the interaction in a medium is distinguishable from those in a vacuum and they possess effective mass even in a fixed state. This can be explained as follows. The vacuum state of the electromagnetic field, when it interacts with the external field of a medium, breaks the symmetry and the transition from this to the "vacuum state" in a medium takes place. In this case, interactions become weaker and the state becomes more stable. According to the Higgs mechanism, this leads to the acquisition of the photon mass. Generally, unlike those in a vacuum, photons in a medium can exist as normal, abnormal or longitudinal photons.

As is well known, the mass of the particle can be divided into two distinct parts:

$$m' = m + \delta m \tag{1}$$

where $m$, $m'$ are masses in the free particle state and in the interaction state respectively and $\delta m$ is a correction due to the interaction. On the other hand, when we perform a renormalization in a perturbative high order approximation, the renormalized propagator function of the photon is

$$D(k, \lambda) \sim \frac{1}{k^2 - \lambda}, \tag{2}$$

where $\lambda$ is a virtual photon rest mass, which is involved in principle in the renormalization process. Physically, this means that photon must possess a rest mass in the form of the effective mass in the perturbative high order approximation.

The propagator function of the photon in the process such as Bhabha scattering, where the established theory of high-energy Quantum Electrodynamics(QED) is not well consistent with the experimental facts is given by

$$D(k, \Lambda_r) \sim \frac{1}{k^2} \pm \frac{1}{k^2 - \Lambda_r}, \tag{3}$$

where $\Lambda_r$ is determined from the comparison with the experiment and corresponds to the effective photon mass. This is related to the fact that the photon, when it is in a high energy state, belongs to the interaction state, not to the free state.

Let us consider the effective photon mass in a medium. According to the superconducting theory, the London equation for the electromagnetic field in a superconducting medium is

$$(\Box + k^2) A_\mu = 0, \tag{4}$$

where $k = (em/2\lambda)^{1/2}$ is a characteristic quantity and corresponds to the photon mass in a superconducting medium. This phenomenon is usually interpreted as a mass acquisition of photons by spontaneous symmetry breaking at low-energy scales and the mass of a superconducting medium is regarded as a Higgs field.

According to the quantum theory of electromagnetic fields in a medium, the propagator function of the photon in a medium reads

$$D(k, m_\lambda) \sim \frac{1}{k^2 - m_\lambda^2}, \tag{5}$$

where $m_\lambda = \omega(1 - n^2)^{1/2}$ with refractive index of a medium $n$ and a frequency $\omega$.

As can be seen, photons in a medium, unlike those in a vacuum, possess rest masses as effective masses. Therefore, in this paper, we call the photon mass $m_\lambda$ the effective mass.

## 2. Spontaneous symmetry breaking and the effective photon mass

Now we apply Lagrangian formulation to the photon mass problem on the basis of spontaneous breaking of Lorentz invariance. Firstly, we use Chern-Simons gauge field theory. Generally, the Lagrangian for Chern-Simons gauge field is (to second order)

$$L = -\frac{1}{4}F_{\mu\nu}F^{\mu\nu} + \frac{k}{2}\varepsilon^{\mu\nu\rho}A_\mu\partial_\nu A_\rho - \frac{kg}{6}\varepsilon^{\mu\nu\rho}\varepsilon^{abc}A_\mu^a A_\nu^b A_\rho^c, \tag{6}$$

where $k$ is a Chern-Simons coefficient, $g$ is a second order expansion coefficient, $A_\mu(x)$ is a field function, $F_{\mu\nu}(x)$ is a field strength tensor and $\varepsilon^{\mu\nu\rho}$ is a Levi-Civita symbol.

Let us assume that Lorentz invariance is spontaneously broken by the dynamical interaction with the medium field and this results in the acquisition of the photon mass in the form of the effective mass.

Taking into account the fact that photons consist of normal, abnormal and longitudinal photons according to polarization states in a medium and their masses are manifested differently, the Lagrangian is

$$L = L_1 + L_2 + L_3, \tag{7}$$

where $L_1$ is a Lagrangian of a pure electromagnetic field and reads

$$L_1 = -\frac{1}{4}F_{\mu\nu}F^{\mu\nu}, \tag{8}$$

$L_2$ is an interaction Lagrangian between a spinor field and the electromagnetic field and reads

$$L_2 = \overline{\psi}(x)[i\gamma^\mu(\partial_\mu + ieA_\mu) - m]\psi(x), \tag{9}$$

and $L_3$ is a Lagrangian which represents the spontaneous breakdown of Lorentz invariance by Chern-Simons gauge field and reads

$$L_3 = \frac{k_i}{2}\varepsilon^{\mu\nu\rho}A_\mu\partial_\nu A_\rho - \frac{k_i g}{6}\varepsilon^{\mu\nu\rho}\varepsilon^{abc}A_\mu^a A_\nu^b A_\rho^c. \tag{10}$$

Chern-Simons coefficients $k_i$ is related to the effective photon masses and $k_1, k_2, k_3$ correspond to normal, abnormal and longitudinal photons in a medium.

Using the action integral

$$S = S_0 + \int d^4x\, \varepsilon^{\mu\nu\lambda\sigma}A_\mu F_{\nu\lambda}k_\sigma \tag{11}$$

and $\delta S = 0$, we can write the field equation(to the first order Chern-Simons term) as

$$\partial_\mu F^{\mu\nu} + 2k_\sigma \varepsilon^{\sigma\nu\rho\beta}F_{\rho\beta} = 0. \tag{12}$$

Comparing the above equation with the covariant form of Maxwell's equations, we can see that the Chern-Simons term $2k_\sigma\varepsilon^{\sigma\nu\rho\beta}F_{\rho\beta}$ is newly added. It is the coefficient $k_\sigma$ of this term that reflects the effective photon mass. $k^2 > 0$, $k^2 < 0$ and $k^2 = 0$ correspond to time-like, space-like and light-like values respectively.

Since three types of photons exist in principle in a medium, the following relations hold.

$$m_1 = 2k_1, \; m_2 = 2k_2, \; m_3 = 2k_3$$

Therefore, when the photon propagates through matter, e.g., interacts with a medium, Chern-Simons theory is naturally explained. It follows that the photons acquire masses owing to the spontaneous breakdown of Lorentz symmetry and they are manifested as effective masses which correspond to three photons in a medium. This is analogous to Coleman-Weinberg effect. When the value of $k_\sigma$ is space-like, it is known that $|k| < 10^{-33}\,\mathrm{eV}$. This indicates that the effective photon mass we are discussing now is less than $10^{-69}\,g$.

### 3. The polarization of a medium and the effective photon mass

Now we deal with the effective photon mass problem in connection with the polarization of a medium and the moving medium. Generally, the Lagrangian in light of the existence of a medium is

$$L = -\frac{1}{4}\left[F_{\mu\nu}F^{\mu\nu} + S_{\mu\nu}S^{\mu\nu} - 2F_{\mu\nu}S^{\mu\nu}\right] \tag{13}$$

where $S_{\mu\nu}$ is a polarization-magnetization tensor. When taking into account the intrinsic electric and magnetic moments of a particle, the Lagrangian can be expressed in terms of four dimensional dipole moment tensor $M_{\mu\nu}$ as

$$L = -\frac{1}{4}\left(F_{\mu\nu}F^{\mu\nu} + S_{\mu\nu}S^{\mu\nu}\right) - \frac{1}{2}\left[F_{\mu\nu}S^{\mu\nu} + M_{\mu\nu}\left(F^{\mu\nu} - S^{\mu\nu}\right)\right]. \tag{14}$$

We can write the Lagrangian in a medium separately in the case when Chern-Simons coefficient is a scalar or vector.

$$\begin{aligned} L_m &= -\frac{1}{4}G_{\mu\nu}G^{\mu\nu} + \frac{1}{4}k_\mu \varepsilon^{\mu\nu\rho}\partial_\nu A_\rho \\ L_m' &= -\frac{1}{4}G_{\mu\nu}G^{\mu\nu} + \frac{1}{2}k_\sigma \varepsilon^{\mu\nu\lambda\sigma} A_\mu G_{\nu\lambda} \end{aligned} \tag{15}$$

When taking into account the polarization-magnetization tensor, these can be expressed more generally as

$$\begin{aligned} L_m &= -\frac{1}{4}\left(F_{\mu\nu}F^{\mu\nu} + S_{\mu\nu}S^{\mu\nu} - 2F_{\mu\nu}S^{\mu\nu}\right) + \frac{1}{4}k\varepsilon^{\mu\nu\rho}\partial_\nu A_\rho \\ L_m' &= -\frac{1}{4}\left(F_{\mu\nu}F^{\mu\nu} + S_{\mu\nu}S^{\mu\nu} - 2F_{\mu\nu}S^{\mu\nu}\right) + \frac{1}{2}k_\sigma \varepsilon^{\mu\nu\lambda\sigma} A_\mu G_{\nu\lambda} \end{aligned} \tag{16}$$

Substituting the above Lagrangians into the Lagrangian equation for the field, corresponding field equations are obtained and the effective photon mass due to the medium effect and the Chern-Simons term is revealed in this case.

Let us consider the effective photon mass problem in a moving medium.

On the basis of Chern-Simon gauge theory, medium equations when the coupling constants are vectors are given by

$$\hat{G}^{\mu\nu} = \Gamma^{\mu\alpha}\Gamma^{\nu\beta}F_{\alpha\beta}$$
$$F_{\alpha\beta} = \partial_\alpha A_\beta - \partial_\beta A_\alpha$$
$$\Gamma^{\mu\alpha} = g^{\mu\alpha} + \lambda u^\mu u^\alpha$$
$$(\lambda = \varepsilon\mu - 1)$$
(17)

where $F_{\alpha\beta}$ is a strength tensor of the electromagnetic field in a vacuum and $\Gamma^{\mu\alpha}$ is a four dimensional medium constant tensor. In general, Chern-Simons Lagrangian in the case of a moving medium can be constructed as

$$L = -\frac{1}{4}\hat{G}_{\mu\nu}\hat{G}^{\mu\nu} + \frac{1}{2}k_\sigma \varepsilon^{\mu\nu\lambda\sigma} A_\mu \hat{G}_{\nu\lambda}.$$
(18)

Therefore, the field equation is given by

$$\partial_\mu \hat{G}^{\mu\nu} + 2k_\sigma \varepsilon^{\sigma\nu\alpha\beta}\hat{G}_{\alpha\beta} = 0.$$
(19)

In this case, the second term in the above equation corresponds to the effective mass of the photon. It can also be described by four dimensional medium constant tensor as follows.

$$\Gamma^{\mu\alpha}\Gamma^{\lambda\beta}\left(\partial_\mu F_{\alpha\beta} + 2k^\sigma \varepsilon_{\sigma\lambda\mu\nu} F_{\alpha\beta}\right) = 0$$
(20)

This equation, in the case of motionless medium and vacuum respectively, leads to

$$\partial_\mu G^{\mu\nu} + 2k_\sigma \varepsilon^{\sigma\nu\alpha\beta} G_{\alpha\beta} = 0$$
$$\partial_\mu F^{\mu\nu} + 2k_\sigma \varepsilon^{\sigma\nu\alpha\beta} F_{\alpha\beta} = 0$$
(21)

The gauge condition for a moving medium is given by

$$\partial_\mu A_\alpha \Gamma^{\mu\alpha} = \partial_\mu A^\mu + \lambda(u_\mu A^\mu)(u_\alpha A^\alpha) = 0.$$

For the moving medium, the following conditions are satisfied in principle.

$$\Gamma^{\mu\alpha}\Gamma^{\lambda\beta} = 0$$
$$\partial_\mu F_{\alpha\beta} + 2k^\sigma \varepsilon_{\sigma\lambda\mu\nu} F_{\alpha\beta} = 0$$
(22)

Actually, the second condition is physically possible.

From the above theoretical reasons, we can conclude that photons have effective rest masses in a moving medium.

### 4. Electric and magnetic charges of particles and the effective photon mass

Generally, the Lagrangian taking into account both the electric and magnetic charges, when the Lorentz symmetry is satisfied is given by

$$L_0 = -\frac{1}{4}\left[F_{\mu\nu}(e)F^{\mu\nu}(e) + F_{\mu\nu}(g)F^{\mu\nu}(g)\right] + j_\mu(e)A^\mu(e) + j_\mu(g)A^\mu(g),$$
(23)

where $F_{\mu\nu}(e)$ and $F_{\mu\nu}(g)$ are tensors of the electromagnetic fields for the electric and the magnetic charges respectively, $j_\mu(e)$ and $j_\mu(g)$ are current densities, $A^\mu(e)$ and $A^\mu(g)$ are vector potentials.

If the photon possessed finite mass, it could be divided into two parts: one associated with the electric charge and the other with the magnetic charge. That is,

$$L_0 = -\frac{1}{4}\left[F_{\mu\nu}(e)F^{\mu\nu}(e) + F_{\mu\nu}(g)F^{\mu\nu}(g)\right] + \frac{1}{2}\left[m_e^2 A_\mu(e)A^\mu(e) + m_g^2 A_\mu(g)A^\mu(g)\right], \tag{24}$$

where $m_e$ and $m_g$ are contributions of electric and magnetic charges to the photon mass, respectively.

If we assume that the photon acquires a mass by the occurrence of the spontaneous breakdown of Lorentz symmetry on the basis of Chern-Simon gauge theory, the Lagrangian is given by

$$L = L_0 + \frac{1}{4}K(e)e^{\mu\nu\rho}A_\mu(e)\partial_\nu A_\rho(e) + \frac{1}{4}K(g)e^{\mu\nu\rho}A_\mu(g)\partial_\nu A_\rho(g), \tag{25}$$

where Chern-Simons coefficients $K(e)$ and $K(g)$ are physical quantities corresponding to electric and magnetic charges of the particle respectively and represent masses.

If $K(g) = 0$ in (25), the result is consistent with the previous research. Generally, in the case when Chern-Simons coefficient is a vector, the Lagrangian is given by

$$L = L_0 + \frac{1}{2}K_\sigma(e)e^{\mu\nu\lambda\sigma}A_\mu(e)F_{\nu\lambda}(e) + \frac{1}{2}K_\sigma(g)e^{\mu\nu\lambda\sigma}A_\mu(g)F_{\nu\lambda}(g), \tag{26}$$

where Chern-Simons coefficients $K_\sigma(e)$ and $K_\sigma(g)$ are four dimensional vectors corresponding to electric and magnetic charges of the particle respectively and indicate $K_\sigma(k_0, k_1, k_2, k_3)$. This is analogous to the fact that the photons reveal the physical properties as normal, abnormal and longitudinal photons in a medium.

In this paper, we construct the motion equations from equations (25) and (26) as follows according to whether Chern-Simons coefficient is a scalar or vector.

1) When Chern-Simons coefficient is a scalar,

$$\begin{aligned}\partial_\mu F^{\mu\nu}(e) + 2K(e)e^{\nu\alpha\beta}F_{\alpha\beta}(e) = 0 \\ \partial_\mu F^{\mu\nu}(g) + 2K(g)e^{\nu\alpha\beta}F_{\alpha\beta}(g) = 0\end{aligned} \tag{27}$$

2) When Chern-Simons coefficient is a vector,

$$\begin{aligned}\partial_\mu F^{\mu\nu}(e) + 2K_\sigma(e)e^{\sigma\nu\alpha\beta}F_{\alpha\beta}(e) = 0 \\ \partial_\mu F^{\mu\nu}(g) + 2K_\sigma(g)e^{\sigma\nu\alpha\beta}F_{\alpha\beta}(g) = 0\end{aligned} \tag{28}$$

As we can see from equations (27) and (28), Chern-Simons coefficient $K$ corresponds to the photon mass e.g., Chern-Simons coefficient $K(e)$ is a mass term associated with the electric charge of the particle and $K(g)$ is the one related to the magnetic charge. Therefore, according to the electromagnetic theory for electric and magnetic charges of the particle, the photon may possess the effective mass which has dual characteristics.

In general, the Lagrangian of the electromagnetic field taking into account electric and magnetic charges of the particle in a medium is given by

$$L' = -\frac{1}{4}\left[G_{\mu\nu}(e)G^{\mu\nu}(e) + G_{\mu\nu}(g)G^{\mu\nu}(g)\right], \qquad (29)$$

where electromagnetic field tensor $G_{\mu\nu}$ is determined by the polarization-magnetization tensor $S_{\mu\nu}$ as follows.

$$G_{\mu\nu} = F_{\mu\nu} - S_{\mu\nu} \qquad (30)$$

The Lagrangian of the electromagnetic field by electric and magnetic charges of the particle can be written as

$$L' = -\frac{1}{4}[F_{\mu\nu}(e)F^{\mu\nu}(e) + F_{\mu\nu}(g)F^{\mu\nu}(g) + \\ + S_{\mu\nu}(e)S^{\mu\nu}(e) + S_{\mu\nu}(g)S^{\mu\nu}(g) - 2(F_{\mu\nu}(e)S^{\mu\nu}(e) + F_{\mu\nu}(g)S^{\mu\nu}(g))] \qquad (31)$$

If we take into account electric and magnetic moments of the particle, the above equation can be written in terms of four dimensional dipole moment tensor $M_{\mu\nu}$ as follows.

$$L' = -\frac{1}{4}[F_{\mu\nu}(e)F^{\mu\nu}(e) + F_{\mu\nu}(g)F^{\mu\nu}(g) - S_{\mu\nu}(e)S^{\mu\nu}(e) - S_{\mu\nu}(g)S^{\mu\nu}(g) \\ -2(F_{\mu\nu}(e)S^{\mu\nu}(e) + F_{\mu\nu}(g)S^{\mu\nu}(g))] + \frac{1}{2}M_{\mu\nu}(F^{\mu\nu}(e) + F^{\mu\nu}(g) - S^{\mu\nu}(e) - S^{\mu\nu}(g)) \qquad (32)$$

It can also be divided into three parts.

$$L' = L_1 + L_2 + L_3$$

where $L_1$ is a Lagrangian in a vacuum due to electric and magnetic charges, $L_2$ is that of the electromagnetic field in a medium due to the polarization-magnetization and $L_3$ is that due to electric and magnetic moments of the particle.

Now the Lagrangian in the case when Chern-Simons coefficient is a scalar or a vector, respectively, is given by

$$L'_1 = -\frac{1}{4}[G_{\mu\nu}(e)G^{\mu\nu}(e) + G_{\mu\nu}(g)G^{\mu\nu}(g)) + \frac{1}{4}Ke^{\mu\nu\rho}(\partial_\nu A_\rho(e) + \partial_\nu A_\rho(g)), \qquad (33)$$

$$L'_2 = -\frac{1}{4}[G_{\mu\nu}(e)G^{\mu\nu}(e) + G_{\mu\nu}(g)G^{\mu\nu}(g)) + \frac{1}{2}Ke^{\mu\nu\lambda\rho}(A_\mu(e)G_{\nu\lambda}(e) + A_\mu(g)G_{\nu\lambda}(g)). \qquad (34)$$

According to the above relations, we can write field equations for two cases.

1) When Chern-Simon coefficient is a scalar,

$$\begin{aligned}\partial_\mu G^{\mu\nu}(e) + 2K(e)e^{\nu\alpha\beta}G_{\alpha\beta}(e) &= 0 \\ \partial_\mu G^{\mu\nu}(g) + 2K(g)e^{\nu\alpha\beta}G_{\alpha\beta}(g) &= 0\end{aligned}. \qquad (35)$$

2) When Chern-Simon coefficient is a vector,

$$\partial_\mu G^{\mu\nu}(e) + 2K_\sigma(e) e^{\sigma\nu\alpha\beta} G_{\alpha\beta}(e) = 0$$
$$\partial_\mu G^{\mu\nu}(g) + 2K_\sigma(g) e^{\sigma\nu\alpha\beta} G_{\alpha\beta}(g) = 0$$ . (36)

From equations (35) and (36), it seems that we just replace $F_{\mu\nu}$ with $G_{\mu\nu}$ when there exists a medium but in reality, the effect of polarization-magnetization tensor $S_{\mu\nu}$ is newly considered. There is an additional field effect caused by the dipole moment tensor $M_{\mu\nu}$ when we consider electric and magnetic charges of the particle and also take into account electric and magnetic dipole moments.

**Conclusion**

We have suggested that although the photon is massless in a free particle state according to the Einstein's theory of relativity, it can possess mass in the interaction states in a medium when particles have electric and magnetic charges.

1. In the perturbative high order approximation of the quantum field theory, the photon can possess mass due to the renormalized photon propagator and it also holds true for the electron-positron Bhabha scattering in high-energy quantum electrodynamics.

2. In a superconducting medium, the photon can acquire a mass according to the London equation for the electromagnetic field.

3. The acquisition of photon mass is also possible in a fixed or a moving medium.

4. In this paper, we showed that the photon has two types of non-vanishing effective masses due to the spontaneous symmetry breaking if the electromagnetic field has magnetic charges as well as electric charges.

According to the above reasons, we conclude that it is possible for the photon to acquire an effective mass on the basis of Chern-Simons gauge theory.

**References**

[1] Y. Hosotani *Phys. Rev.* D 2022 **51** (1995)
[2] V. P. Gusynin *Phys. Rev.* D **52** 4747 (1995)
[3] M. C. Diamantini, G. Guarnaccia and C. A. Trugenberger *arXiv: 1310.2103 [hep-th]* (2013)
[4] L. C. Garcia de Andrade arXiv: *1112.4927 [astro-ph.CO]* (2011)
[5] J. Heeck *arXiv: 1304.2821 [hep-ph]* (2013)
[6] J. Luo, L. –C. Tu, Z. K. Hu and E. –J. Luan E *Phys. Rev. Lett.* A **90** 0818011 (2003)
[7] L. –C. Tu, J. Luo and G. T. Gillies *Rep. Prog. Phys.* **68** 77 (2005)
[8] A. S. Goldhaber and M. M. Nieto *Rev. Mod. Phys.* **82** 939 (2010)

[9] G. Spavieri and G. T. Gillies *arXiv: 1005.4082 [quant-ph]* (2010)

[10] A. Accioly et al *arXiv: 1012.2717 [hep-th]* (2010)

[11] B. G. Sidharth *Found. Phys. Lett* **19(1)** 87 (2006)